\def\year{2021}\relax
\documentclass[letterpaper]{article} 
\usepackage{aaai21}  
\usepackage{times}  
\usepackage{helvet} 
\usepackage{courier}  
\usepackage[hyphens]{url}  
\usepackage{graphicx} 
\usepackage{natbib}  
\usepackage{caption} 
\usepackage{booktabs}
\usepackage{multirow}
\usepackage{multicol}
\usepackage{comment}
\usepackage{xcolor}
\usepackage{dirtytalk}
\usepackage{tabularx}
\usepackage{rotating}
\usepackage{tablefootnote}
\usepackage{hhline}
\usepackage{pbox}
\usepackage{graphicx}
\usepackage{soul}

\renewcommand\hl[1]{#1}

\title{Pathways through Conspiracy: The Evolution of Conspiracy Radicalization through Engagement in Online Conspiracy Discussions}
 \author{
 }

\author {
    Shruti Phadke\textsuperscript{\rm 1},
    Mattia Samory\textsuperscript{\rm 2},
    Tanushree Mitra\textsuperscript{\rm 1} \\
}
\affiliations {
    \textsuperscript{\rm 1} University of Washington, USA \\
    \textsuperscript{\rm 2} GESIS, Germany \\
    phadke@uw.edu, mattia.samory@gesis.org , tmitra@uw.edu 
}

\begin{document}

\maketitle


\begin{flushleft}
      {\color{red} To cite: Shruti Phadke, Mattia Samory,  Tanushree Mitra. 2022. Pathways through Conspiracy: The Evolution of Conspiracy Radicalization through Engagement in Online Conspiracy Discussions. Proceedings of the International AAAI Conference on Web and Social Media (ICWSM) 2022. (accepted March 2022).}
      \vspace{10pt}
\end{flushleft}

\begin{abstract}
The disruptive offline mobilization of participants in online conspiracy theory (CT) discussions has highlighted the importance of understanding how online users may form radicalized conspiracy beliefs. While prior work researched the factors \emph{leading up to} joining online CT discussions and provided theories of how conspiracy beliefs form, we have little understanding of how conspiracy radicalization evolves \emph{after} users join CT discussion communities. 
In this paper, we provide the empirical modeling of various radicalization phases in online CT discussion participants.
To unpack how conspiracy engagement is related to radicalization, we first characterize the users' journey through CT discussions via conspiracy engagement pathways. 
Specifically, by studying 36K Reddit users through their 169M contributions, we uncover four distinct pathways of conspiracy engagement: \textit{steady high}, \textit{increasing}, \textit{decreasing} and \textit{steady low}.
We further model three successive stages of radicalization guided by prior theoretical works.
Specific sub-populations of users, namely those on \textit{steady high} and \textit{increasing} conspiracy engagement pathways, progress successively through various radicalization stages. In contrast, users on the \textit{decreasing} engagement pathway show distinct behavior: they limit their CT discussions to specialized topics, participate in diverse discussion groups, and show reduced conformity with conspiracy subreddits. 
By examining users who disengage from online CT discussions, this paper provides promising insights about conspiracy recovery process.

\end{abstract}

\section{Introduction}

Disinformative, panic-inducing online conspiracy theories (CT) are increasingly proving to be threats to productive civic discourse. Violent riots at the U.S. Capitol \cite{Capitolr71online} and COVID-19 vaccine skepticism \cite{Vaccinec48online} are just two of the many examples of how online CT discussions result in socially harmful situations. Further, by mainstreaming the fringe, social media allow previously disconnected conspiracy theorists to find like-minded individuals, reinforce their beliefs, and mobilize \cite{Marwick2017MediaOnline}. 
Despite the clear implications of online CT engagement, we know little about an individual's journey through CT discussion communities and how they become increasingly engaged (or not) with conspiratorial worldviews. This paper provides just such an understanding. 

\subsubsection{\textbf{What are the pathways of online conspiracy engagement?} }
Using digital traces of 36K Reddit users who participated in {\small \tt r/conspiracy}---the biggest CT discussion subreddit, we characterize their Reddit trajectories \emph{after} they made their first comment in {\small \tt r/conspiracy}. 
Working with their 169M contributions spread across 4K subreddits, we find four distinct types of engagement trends in CT discussion subreddits---\textit{steady high}, \textit{increasing}, \textit{decreasing} and \textit{steady low}. For instance, users with \textit{increasing} engagement contribute consistently and predominantly more over time in CT subreddits compared to other subreddits. 

\subsubsection{\textbf{How does conspiracy radicalization process evolve for users?}}
To answer, we leverage a theoretical model of internet mediated radicalization \cite{Neo2016AnRadicalisation} comprising five phases---\textbf{R}eflection, \textbf{E}xploration, \textbf{C}onnection, \textbf{R}esolution, \textbf{O}peration (RECRO). We focus on the first three phases that are most visible in online discussions \cite{van2019echo}. They describe the psychological and emotional vulnerabilities (\textbf{R}eflection), development of alternative worldviews (\textbf{E}xploration) and social bonds (\textbf{C}onnection) formed during the radicalization process. 
We model the phases through various linguistic, interaction and activity features. For example, 
we characterize the \textbf{E}xploration phase by creating a \textit{generality scale} that models the users' monological conspiracy worldview \cite{goertzel1994belief} signaling adoption of generalized conspiracy thinking.

We find that users with \textit{steady high} and \textit{increasing} engagement in CT do show signs of radicalization through increasing use of insider language, and repeated participation in small-group discussions. Conversely, users with \textit{decreasing} engagement in CT communicate in diverse discussion groups, and never develop lexical conformity with conspiracy communities. Moreover, users with \textit{steady high} and \textit{increasing} engagement, increasingly engage in generalist CT discussion subreddits, showing support for the monologicality hypothesis of conspiracy belief evolution---in stark contrast to users on decreasing pathways who 
limit their contributions to specific CT discussion topics. 

\subsubsection{Contributions.} 
Through a theory-driven, empirical study of the conspiracy radicalization process, our work lays the foundation for observing users who may act on their radicalized beliefs, i.e., those who may proceed to the \textbf{O}peration phase. 
We also offer a \textit{generality scale} used in characterizing conspiracy worldviews, that captures the generalist or specialist nature of subreddits. 
Overall, our work characterizes how conspiracy beliefs evolve during an individual's online lifespan since their first contribution to CT discussions. By differentiating between users with high engagement from those who disengage with CT discussions our work has implications for understanding factors in recovery from online CT discussions.

\section{Related Work}

\subsection{\textbf{Online Conspiracy Engagement}}
People turn to conspiracy beliefs to fulfill their epistemic, existential, and social needs \cite{douglas2017psychology}. 
For example, users who join CT discussion communities during crisis events---characterized by existential threats and uncertainty---engage more and more exclusively with them \cite{samory2018conspiracies}. Further, social activity, such as facing marginalization from mainstream discussion spaces and bonding with incumbent members, correlates with a higher likelihood of joining CT communities \cite{phadke2021makes}. \hl{Scholars investigating the pathways into conspiracy theory engagement found the evidence of self-selection and shared interests that feed into engagement in CT discussions \mbox{\cite{klein2019pathways}}.} Conversely, extreme cognitive dissonance due to conflicting beliefs, correlates with lowered engagement and shorter tenure \cite{phadke2021characterizing}. Similarly, encouragement by trusted peers and exposure to evidence-based counter-narratives prompt people to exit their conspiracy beliefs \cite{xiao2021sensemaking}. 
Adding to this recent body of literature, which uncovers how individuals begin and end their engagement with CT communities, this work connects the unexplored transition between these extremes. Specifically, we investigate how different conspiracy engagement pathways are associated with various phases of internet-mediated radicalization.




\subsection{\textbf{Conspiracy Theorizing and Radicalization}}
By radicalization, we mean the growing support by individuals or groups for a radical societal change, either through belief or action, that harms the social fabric and functioning \cite{dalgaard2010violent}. 
Researchers have identified three broad ways in which conspiracy theorizing can play a role in advancing radicalization. First, conspiracy theorizing can provide a paranoid interpretation of reality in which there are clear in-groups and out-groups that overall enhance the appeal of extremist narratives \cite{vermeule2009conspiracy}. Second, conspiracy ideation and radicalization are believed to contain similar underlying psychological disposition such as feelings of anger, anxiety, paranoia \cite{butter2020routledge}. Third, apart from creating clear boundaries between in-group and out-group, conspiracy theorizing can strengthen in-group bonding which is essential for radicalization process \cite{conway2012zarqawi}. 

Although this literature shows that conspiracy theorizing may lead to radicalization, there is little to no empirical investigation connecting the two phenomena \cite{butter2020routledge}. Especially in light of the recent events where online CT discussions led to riots causing national security issues \cite{Capitolr71online}, we believe that it is important to study if, and how, online CT discussion participants display markers of radicalization. In this paper, we study the process of conspiracy radicalization using a RECRO model for internet-mediated radicalization \cite{Neo2016AnRadicalisation} which we briefly explain next. 




\subsection{Models of Radicalization}
Radicalization models describe the progression of the radicalization process where individuals move from socially normative perspectives towards more extremes ones \cite{neodetecting}. However, extant radicalization pathway models and theories either do not consider the role of internet in radicalization or focus mainly on the psychological predispositions of people \cite{Neo2016AnRadicalisation}. The RECRO model proposed by \citeauthor{Neo2016AnRadicalisation}, however, is a pathway-based theoretical model that views radicalization as an internet-mediated process involving all, individual, epistemic and social factors. Researchers have used RECRO in qualitative analyses of online anti-vaccination discussions finding that social media provides a strong platform for the first three phases \cite{van2019echo}---\textbf{R}eflection, \textbf{E}xploration, \textbf{C}onnection. Hence, we study the first three phases of radicalization in the CT engagement which are briefly explained below.

\subsubsection{\textbf{Reflection: }}
The \textbf{R}eflection phase details the vulnerabilities and psychological predispositions that increase one's receptivity towards radicalization \cite{Neo2016AnRadicalisation}. This is a phase where personality and psychological factors motivate the individuals to open-up, also described as ``cognitive opening'', to alternate belief systems. Other researchers also agree on the importance of psychological footprints such as anger and heightened emotions in online radicalization \cite{dalgaard2010violent}. After the cognitive opening, users begin to form radical worldviews in the \textbf{E}xploration phase.


\subsubsection{\textbf{Exploration: }} Here, individuals begin to make sense of new information and narratives by forming alternate worldviews in a way that fosters eventual radicalization \cite{Neo2016AnRadicalisation}. 
\hl{Individuals are primed to form a new, alternate worldview that resonates with their interests and epistemological needs \mbox{\cite{Neo2016AnRadicalisation}}}. 
Specifically in relation to conspiracy theorizing, researchers propose a ``monological belief system,'' describing it as a stable cognitive style that dictates the perceived 
functioning of the social world \cite{goertzel1994belief}. 
Monological conspiracy worldviews offer a general set of assumptions, such as cover-ups by powerful people, that are portable across multiple CTs and socio-political phenomena, independently from their specific topic or context \cite{goertzel1994belief}. 
\hl{This affords applying CT to any socio-political phenomena, independently from the specific topic or context of an event \mbox{\cite{Franks2017BeyondWorldviews}}.} However, 
%
Hence, the monologicality hypothesis paints the picture of a closed-minded CT believer with a strong mobilization potential, and of a CT ecosystem of broadly applicable, interconnected, mutually supporting ideas. This hypothesis though is contested. Competing research presents a possibility of better educated, open, and socially active CT believers who might restrict their interests to specific conspiracy topics \cite{franks2017beyond}. This paper, for the first time, analyzes the generality or specificity of CT belief by modeling how individuals \emph{explore} the world of online conspiracies after their initial exposure. 

\begin{figure*}[t]
    \centering
    \includegraphics[width=0.99\textwidth]{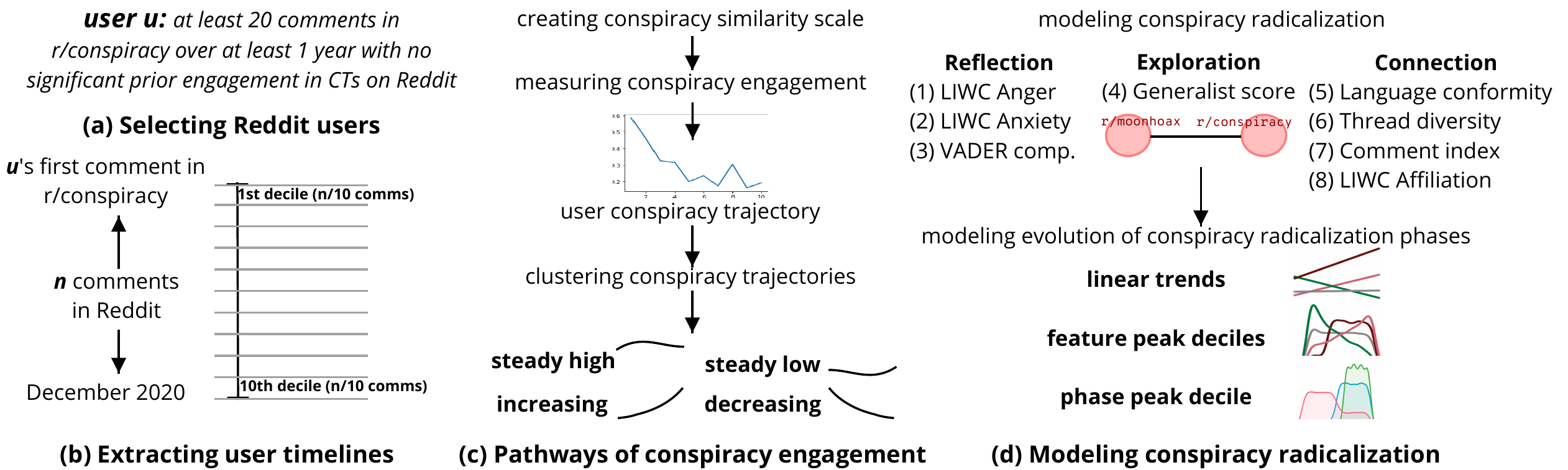}
    \caption{(a) \&(b) describe user selection and data preparation processes. We split the user activity into 10 deciles of equal contribution volumes after the user's first comment in {\small \tt r/conspiracy} as described in ``Data'' Section. (c) We then model the conspiracy engagement pathways by unsupervised clustering of user trajectories ``Pathways of Conspiracy Engagement'' Section. (d) Finally, we model the conspiracy radicalization process by operationalizing the first three phases in the RECRO model \cite{Neo2016AnRadicalisation} through 8 features in ``Characterizing Conspiracy Radicalization'' Section.}
    \label{fig:metafig}
\end{figure*}

\subsubsection{\textbf{Connection: }} 
Here, individuals interact to form group bonds with like-minded people \cite{Neo2016AnRadicalisation}. As opposed to the \textbf{R}eflection phase capturing individual predisposition, the \textbf{C}onnection phase describes how bonds with a group of peers advance the radicalization process. Specifically, cohesion or conformity to one's social group \cite{crossett2010radicalization}, small-group dynamics \cite{reedy2013terrorism} and feelings of group affiliation \cite{dalgaard2010violent} are strongly associated with radicalization. 

\section{Data}
\label{sec:data}

\subsubsection{\textbf{Selecting Reddit users: }} In this paper, we study the evolution of conspiracy radicalization in social media users \emph{after} their initial engagement in CT discussions on Reddit. To mark the users' entry into Reddit's CT discussion world, we look at users' first comment into {\small \tt r/conspiracy}---Reddit's biggest and most popular conspiracy discussion community. Our goal is to analyze users' long-term engagement in online CT discussions. Hence, we select 42,225 users that contribute at least 20 comments in {\small \tt r/conspiracy} over at least one year. Having such activity thresholds ensures that our analysis is not biased by users that have sparse involvement in CT discussions, a practice commonly followed in the social computing research \cite{kumar2018community,samory2018conspiracies}. Additionally, we filter the user list based on other robustness checks detailed in Section ``Additional Robustness Checks'', to avoid selection bias and ensure fair activity coverage across all users. 
Finally, the filtered dataset includes 36,314 users. 

\subsubsection{\textbf{Extracting user timelines: }} Reddit users differ in their frequency and levels of contribution. To compare evolution of all users on equal footing, we split their Reddit activity in ten equal deciles of contribution volume. For example, if a user $x$ makes their first comment in {\small \tt r/conspiracy} on 1st January 2018 and makes $n$ comments on Reddit since then, we split their entire Reddit activity \emph{after} 1/1/2018 into equal, time ordered batches of $n/10$ comments (Fig. \ref{fig:metafig} (b)). 
This approach ensures that (1) we have ten deciles worth of activity with an equal number of contributions in each decile for every user and (2) that each user is evaluated according to their own pace of engagement in CT discussions. A similar approach for mitigating temporal shifts in user activity has been validated in studying far-right radicalization on Twitter \cite{vidgen2021islamophobes}. 

Thus, we collect the entire activity for 36,314 users after their first post in {\small \tt r/conspiracy} and prior to December 2020, split over 10 deciles of equal contribution volumes. \hl{In order to eliminate subreddits that see contributions from users only occasionally, we kept subreddits that had at least 10 contributions from at least 5 different authors, spreading the user activity over 4,756 subreddits and 169M comments.} We mine all data using Pushshift \cite{baumgartner2020pushshift}.

\section{Pathways of Conspiracy Engagement}
\label{sec:pathway_methods}
To understand a user's journey through Reddit after participating in conspiracy discussions, we first model each user's longitudinal development of CT engagement. Fig. \ref{fig:metafig} (c) outlines our process for extracting pathways of conspiracy engagement. 
We start by describing the \textit{conspiracy similarity scale} and our methods for validating the scale.

\begin{figure*}[t]
    \centering
    \includegraphics[width=0.99\textwidth]{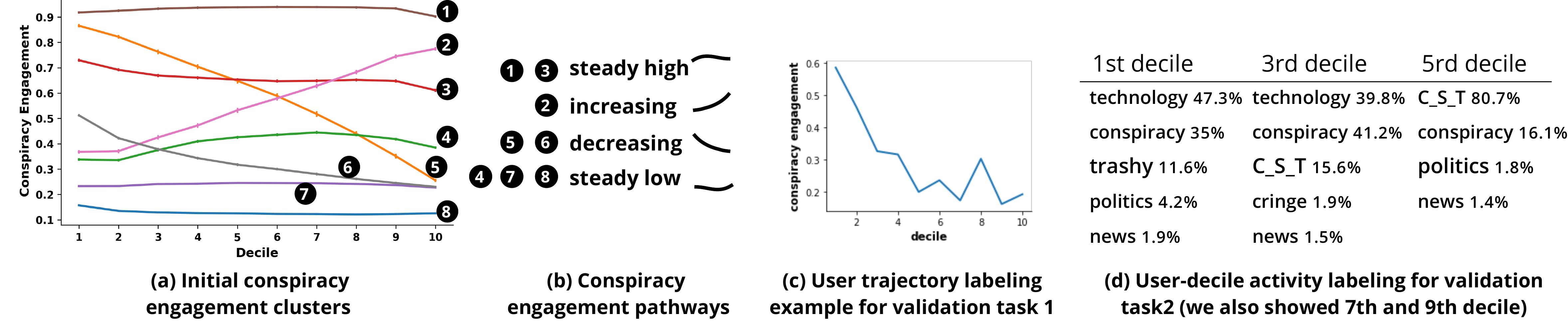}
    \caption{ (a) Plot displaying average trajectories across 8 conspiracy engagement clusters. Here we can observe that some trajectories show similar trend, only shifted in amplitude. For example, both 5 \& 6 show decreasing engagement. Hence we group together amplitude shifted trajectories as displayed in (b). These are the final conspiracy engagement pathways analyzed throughout this paper. (c) Example user trajectory plots shown to evaluators for validation task 1. (d) Example user-decile activity shown to the evaluators for validation task 2.}
    \label{fig:kmeans_results}
\end{figure*}

\subsection{Creating the Conspiracy Similarity Scale}
To understand how conspiratorial are the discussions in each subreddit, we use a scalar \textit{conspiracy similarity scale} inspired by the methods described by \citet{samory2018conspiracies}. Specifically, we map subreddits on a scale from -1 to 1 where 1 represents the highest similarity to {\small \tt r/conspiracy}.
\hl{Previous works find polarized communities in CT and scientific news consumption patterns \mbox{\cite{Bessi2015ScienceMisinformation}}. Moreover, in terms of psychology and norms, there are known biases associated with conspiracy beliefs that would be unacceptable in scientific communities \mbox{\cite{kuhn2021coronavirus}}. Hence, to find the CT discussion communities, we contrast the user activity in {\small \tt r/conspiracy} with its polar opposite and the largest scientific community---{\small \tt r/science}.}
Using the Reddit activity of users with at least 10 comments in {\small \tt r/conspiracy} or {\small \tt r/science}, we create a vector representation for each subreddit by calculating pointwise mutual information (PMI) between each pair of subreddits. 
\hl{More simply put, PMI is a co-occurrence based measure \mbox{\cite{bouma2009normalized}} that can characterize similarity between two subreddits based on the number of commonly occurring users in them. PMI provides a high dimensional matrix where the pointwise mutual information is provided for each pair of subreddits in the dataset. Hence, to create a lower dimensional embedding for each subreddit, we calculate the singular value decomposition (SVD) matrix on PMI. 
Finally, to characterize a subreddit's similarity to CT discussions, we calculate the cosine similarity of each SVD vector with the vector for {\small \tt r/conspiracy}}. The final \textit{conspiracy similarity scale} has scores for 4,756 subreddits with top most subreddits similar to {\small \tt r/conspiracy} listed in Table \ref{tab:conspiracy_sim_examples}.


\subsubsection{\textbf{Validating the \textit{ Conspiracy Similarity Scale}: }} To validate the \textit{conspiracy similarity scale} 
we refer to the concept of convergent validity
that measures the correlation between the \textit{conspiracy similarity scale} with other subreddit similarity measures based on the same construct. For this comparison we use the only publicly available subreddit embeddings contributed by  \citet{kumar2018community}.
While those subreddit embeddings are not specifically catered towards finding CT subreddits, the embeddings do provide a general measure of subreddit similarity. 
Specifically, we calculate Spearman's rank-order correlation between the 1000 subreddits most similar to {\small \tt r/conspiracy} according to our and Kumar's rankings. We find a significant ($p < 0.05$) moderate correlation ($0.52$) between the two. This corroborates that our \textit{conspiracy similarity scale} successfully measures similarity to {\small \tt r/conspiracy}. We further manually analyze the top 100 subreddits on the \textit{conspiracy similarity scale} and confirm that they do host CT discussions. 

\begin{table}[]
\centering
\resizebox{0.7\linewidth}{!}{%
\begin{tabular}{ll}
{\small \tt conspiracy}     & {\small \tt ConspiracyMemes}  \\
{\small \tt The\_Donald}    & {\small \tt conspiracyundone} \\
{\small \tt C\_S\_T}        & {\small \tt ConspiracyII}     \\
{\small \tt UFOs}           & {\small \tt occult}           \\
{\small \tt greatawakening} & {\small \tt AskThe\_Donald}   \\ \hline
\end{tabular}%
}
\caption{Top 10 subreddits on conspiracy similarity scale.}
\label{tab:conspiracy_sim_examples}
\end{table}



\subsection{Measuring Conspiracy Engagement}

\subsubsection{\textbf{Conspiracy engagement measure: }} The conspiracy engagement measure captures what proportion of the user's Reddit activity is dedicated to CT discussion subreddits in each decile. 
We calculate the weighted average of contributions in each subreddit weighted by that subreddit's score on the \textit{conspiracy similarity scale}. More formally, for a user $u$, with $N$ contributions in a decile $i$, their conspiracy engagement score $C_{u}^i$ will be calculated as:
\(
    C_{u}^i = \sum_{j=1}^{J} n_{j} s_{j}/N,
\)
where $n_{j}$ is number of contributions in the $j_{th}$ subreddit and $s_{j}$ is the subreddit's conspiracy similarity score. $C_{u}^i$ is bounded between 0 to 1 with higher scores indicating higher proportion of engagement in the CT discussion subreddits. 

\subsection{\textbf{Clustering Conspiracy Trajectories}}
We represent every user's conspiracy engagement trajectory as a time series of $C_{u}^i$ over ten deciles. We perform unsupervised clustering of the user trajectories to find common temporal patterns, or pathways, of conspiracy engagement---a method commonly used to characterize longitudinal behavior \cite{genolini2016kmlshape}. 
We next discuss our choice of clustering algorithm, distance measure, and number of clusters. 

In a large-scale benchmark spanning over 128 synthetic, natural and pre-processed datasets, \citet{javed2020benchmark} found that there is no particular algorithm that outperforms others in time series clustering. However, the dynamic time warping (DTW) distance measure outperformed alternatives for a nontrivial number of datasets. DTW is especially suited for time series clustering for its robustness to time shifts between different trajectories \cite{javed2020benchmark}. We thus combine K-Means clustering with a DTW distance measure. We make this choice to complement the computational complexity of DTW with fast, adaptable and convergent cluster assignments produced by K-Means. 
We tune the number of clusters by training K-Means models for $k \in \{2, \ldots, 15\}$, and by identifying the elbow point in the silhouette coefficients. We select $k=8$ as a trade-off between number of clusters and cluster quality. 




\subsection{\textbf{Results: Conspiracy Engagement Pathways }}
Fig. \ref{fig:kmeans_results} (a) shows the results from our cluster analysis. We observe distinctive increasing and decreasing conspiracy engagement patterns across the 8 clusters. While DTW is not sensitive to time (x-axis) shifts in the trajectories, it is still sensitive to magnitude (y-axis) shifts. For example, clusters 5 and 6 in Fig. \ref{fig:kmeans_results}~(a) both show decreasing trend but are shifted in magnitude. Hence, we further group the magnitude-shifted clusters based on common patterns. Finally, we end up with four distinct patterns of engagement representing groups of trajectories---\textit{steady high}, \textit{increasing}, \textit{decreasing} and \textit{steady low}---as described in Fig. \ref{fig:kmeans_results} (b).  

\subsubsection{\textbf{Validating conspiracy engagement pathways: }} 
We invited 6 evaluators proficient in statistics and data analysis to manually assess the quality of conspiracy engagement pathway assignments in two annotation tasks.


\noindent \textbf{1. User trajectory labeling:} We asked the evaluators to label a user conspiracy trajectory plot---for example, the trajectory displayed in Fig. \ref{fig:kmeans_results} (c)---as either \textit{steady high}, \textit{increasing}, \textit{decreasing} or \textit{steady low}. As instructions, we additionally provided $C_{u}$ thresholds for each pathway and demonstrated sample trajectories from each pathway as a guideline for annotations.  

\noindent \textbf{2. User-decile activity labeling:} 
We showed user contributions over subreddits in every other (1st, 3rd, 5th, 7th, 9th) decile (Fig. \ref{fig:kmeans_results} (d)) and asked the evaluators to label the user activity by one of the four conspiracy pathways.

We randomly selected 12 users from each conspiracy pathways separately for each of the tasks. Each trajectory was labeled by two evaluators. We consider a true positive assessment for a trajectory only when \emph{both} evaluators agree on the conspiracy pathway label. Evaluators labeled trajectories in task 1 with an accuracy of 78\%, while 83\% validation samples received perfect agreement. 
Task 2 resulted in accuracy of 84\% accuracy with a perfect agreement of 96\%. 
The validation performance across both tasks suggests that the computational conspiracy pathway assignments are cohesive and can be inferred through both, user trajectory plot (task 1) and the user's raw activity data (task 2). 

\section{Characterizing Conspiracy Radicalization}
\label{sec:rad_methods}
Next, we model the conspiracy radicalization process for users along the four conspiracy engagement pathways. We leverage the radicalization framework outlined in the RECRO model and operationalize the first three phases. Fig. \ref{fig:metafig} (d) displays the summary of all features. 

\subsection{Characterizing Reflection Phase}
The \textbf{R}eflection phase captures the psychological predispositions of users towards adopting radicalization narratives online \cite{Neo2016AnRadicalisation}. Predisposition towards radicalization can be visible through the psycho-linguistic footprints left by the users online \cite{dalgaard2010violent}. Specifically, researchers found that language reflecting anger, anxiety and heightened emotions was used by online radicalized groups \cite{dalgaard2010violent}. 


Hence, to measure the language related to anger and anxiety, we use \textbf{anger} and \textbf{anxiety} lexicons, respectively, from Linguistic Inquiry and Word Count (LIWC) dictionary \cite{tausczik2010psychological}. LIWC encodes words capturing affective, emotional and cognitive processing expressions and is often used for psycho-linguistic analysis of online texts. To measure emotionality, we calculate the average \textbf{compound} VADER sentiment scores \cite{hutto2014vader}. In total, we calculate 3 linguistic features to characterize the \textbf{R}eflection phase.

\subsection{Characterizing Exploration Phase}
The \textbf{E}xploration phase describes a period in which users develop alternate worldviews that advance the radicalization process. Specifically in conspiracy theorizing, scholars have debated whether conspiracy theory belief evolves into a monological worldview---a tendency to analyze all events through the lens of conspiracy theorizing \cite{goertzel1994belief}. \hl{Previous researchers have concluded that online discussions could be useful in understanding the users' conspiracy worldview \mbox{\cite{wood2015online}.}}
To understand how online users explore the world of Reddit CTs, here we characterize conspiracy worldviews by calculating conspiracy generalist or specialist engagement. Specifically, we create a \textit{generality scale}, that scores a subreddit based on the generality of topic discussions.


\subsubsection{\textbf{Creating generality scale}}
\hl{One of the prominent perspectives in research on conspiracy world-views focuses on the ``monological belief system'' 
\mbox{\cite{goertzel1994belief}}. The  monological  perspective  describes  conspiracy belief as closed in itself in which, each conspiracy belief reinforces another. That is, belief in one CT is correlated with belief in other conspiracy theories \mbox{\cite{Swami2011ConspiracistTheories}}. Such generalized conspiracy thinking stemming out of monological worldview could contribute towards more extreme belief in conspiracies. 
In subreddit generality scale, we want to capture the extent of generality (or specificity) of topics discussed in any subreddit. 
For example {\small \tt r/conspiracy} hosts more general CT discussions compared to {\small \tt r/moonhoax} which focuses on a specific moon landing conspiracy. Reddit houses thousands of such general and special discussion subreddits that allow users to participate in a topic with different levels of specialization. How can we computationally determine how generalist a subreddit is?}

\hl{Previous researchers have proposed a generalist-specialist ranking for subreddits based on how generalist or specialist is the subreddit's user base \mbox{\cite{waller2019generalists}}. However, this scale does not consider the actual topic of discussion through the content in subreddit posts. In this paper, we instead build a subreddit-entity network that simultaneously captures the content discussed in a subreddit and how general that content is across Reddit. }

\begin{figure}[t]
    \centering
    \includegraphics[width=0.65\linewidth]{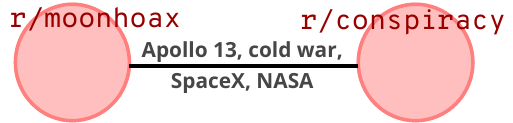}
    \caption{Figure showing example edge connections made in subreddit entity network. We connect two subreddits with an edge if they share same entities (e.g, {\small \tt Apollo 13, cold war}) in top scoring submissions. The aggregate edge weight between two subreddits is the sum of inverse frequencies of shared entities across the entire corpus.}
    \label{fig:entnet}
\end{figure}

\subsubsection{\textbf{Subreddit-entity network: }}
Intuitively, more general subreddits will host content that is less \emph{exclusive} across Reddit. For example, {\small \tt r/conspiracy} hosts political conspiracies on topics also discussed in {\small \tt r/politics} and medical conspiracies on topics in {\small \tt r/science}. However, subreddits such as {\small \tt r/moonhoax} host specific conspiracy topics, that are less likely to be popular across rest of the Reddit. 
We leverage this intuition and build ``subreddit-entity network''. 

Subreddit-entity network is a graph in which subreddits are the nodes connected based on common content. To find the content representative of a subreddit \cite{horne2017identifying}, we analyze the top 200 submissions made in every subreddit and extract named entities (names of people, places, organizations etc.) from the submission text.
We create an edge between two subreddits if top posts in both subreddits mention the same entity (see Fig. \ref{fig:entnet}). 
To further improve the quality of edges, we consider the inverse term frequency of the shared entity across the entire corpus. 
The total edge weight between two subreddits is the sum of inverse term frequency of the entities shared between the subreddits. 

In the subreddit-entity network, intuitively, more generalist subreddits will share more entities with other subreddits. In other words, generalist subreddits might be in a densely connected neighborhood connecting various sub-graphs. Accordingly, generalist subreddits may be \emph{influential} in the subreddit-entity network. Hence, to assess the degree of generality of a subreddit, we calculate the eigenvector centrality for all subreddits 
which is a measure of influence, where a node is considered to be influential if it is connected to other influential nodes. We consider the eigenvector centrality as the subreddit generality score where subreddits with higher eigenvector centrality hosting more general discussions. 

\begin{table}[]
\centering
\resizebox{0.79\linewidth}{!}{%
\begin{tabular}{@{}ll@{}}
\toprule
\multicolumn{1}{c}{\textbf{Most general}} & \multicolumn{1}{c}{\textbf{Most specialist}} \\ \midrule
{\small \tt r/C\_S\_T}                          & {\small \tt r/SacredGeometry}                      \\
{\small \tt r/HighStrangeness}                  & {\small \tt r/flatearth}                           \\
{\small \tt r/conspiracyundone}                 & {\small \tt r/theworldisflat}                      \\
{\small \tt r/conspiracy}                       & {\small \tt r/AntarcticAnomalies}                  \\
{\small \tt r/conspiracytheories}               & {\small \tt r/chemtrails}                          \\ \bottomrule
\end{tabular}%
}
\caption{Most general and most specialist conspiracy discussion subreddits on the subreddit generality scale.}
\label{tab:genscale_val1}
\end{table}

\subsubsection{\textbf{Validating subreddit generality scale: }}
Do eigencentrality values actually represent how generalist or specialist the subreddits are? We validate this in two different steps. 

\noindent \textbf{Step 1:} We created the \textit{subreddit generality scale} to assess the generality of discussions in CT subreddits. Hence, using the \textit{conspiracy similarity scale}, we compile a list of 171 CT discussion subreddits and contrast the generality of their themes with their ranking in the \textit{subreddit generality scale}. 
We manually examine the relative ranking of 171 CT discussion subreddits and validate that the \textit{subreddit generality scale} places generalist CT subreddits on higher end and specialist subreddits on the lower end (example Table \ref{tab:genscale_val1}).

\noindent \textbf{Step 2:} For non CT related subreddits, we create 400 pairs where the first subreddit in the pair is more generalist compared to the second ({\small \tt r/Guitar}  $\rightarrow$ {\small \tt r/AcousticGuitar}). Subreddits included in the 400 pairs range over a diverse list of 45 topics referred from {\small \tt r/ListOfSubreddits}.
We calculate the number of pairs for which our scale scores first subreddit higher than the second finding that 81\% of the pairs are ranked correctly. This indicates that our scale is able to capture the overall generalist-specialist themes in subreddits.

\subsubsection{Generalist engagement scores}
With the \textit{generality scale} we can now measure whether users increasingly engage in more generalist or more specialist CT discussion subreddits. 
Using the same computation as the conspiracy engagement score, we calculate the \textbf{generalist engagement} score in every decile as the weighted average of a user's contributions in subreddits weighted by the subreddits' generality score. Higher generalist score would indicate high engagement in subreddits with general discussions. 


\begin{table}[]
\centering
\resizebox{0.75\linewidth}{!}{%
\begin{tabular}{@{}ccc@{}}
\toprule
\textbf{r/chemtrails} & \textbf{r/AncientAliens} & \textbf{r/McDonalds} \\ \midrule
contrail              & sitchin                  & mcdouble             \\
chemtrail             & anunnaki                 & mcchicken            \\
geoengineering        & nibiru                   & frappe               \\
stratospheric         & panspermia               & mcflurry             \\
haarp                 & gobekli                  & mcnugget             \\ \bottomrule
\end{tabular}%
}
\caption{Example of top 5 words (out of 1000) in SAGE lexicons with first two columns showing CT discussion subreddits. We see that our lexicon captures vocabulary that is distinctively specific to the subreddit. For example, `haarp'' in r/chemtrails refer to H.A.A.R.P project by U.S. Air Force that is theorized to be a weather control weapon. Similarly, 
anunnaki in r/AncientAliens is believed to be a race of ancient aliens. Note that outside of the CT groups, these words have little meaning, making them especially relevant for measuring group language conformity.}
\label{tab:lexconform}
\end{table}

\subsection{Characterizing Connection Phase}
In the \textbf{C}onnection phase, individuals form social connections to support and reinforce their alternate worldviews, facilitating the relational bond between an individual and a wider radical movement \cite{Neo2016AnRadicalisation}. To characterize the this phase, we capture language conformity and group connections established by users with conspiracy communities.

\begin{figure*}[t]
    \centering
    \includegraphics[width=0.99\textwidth]{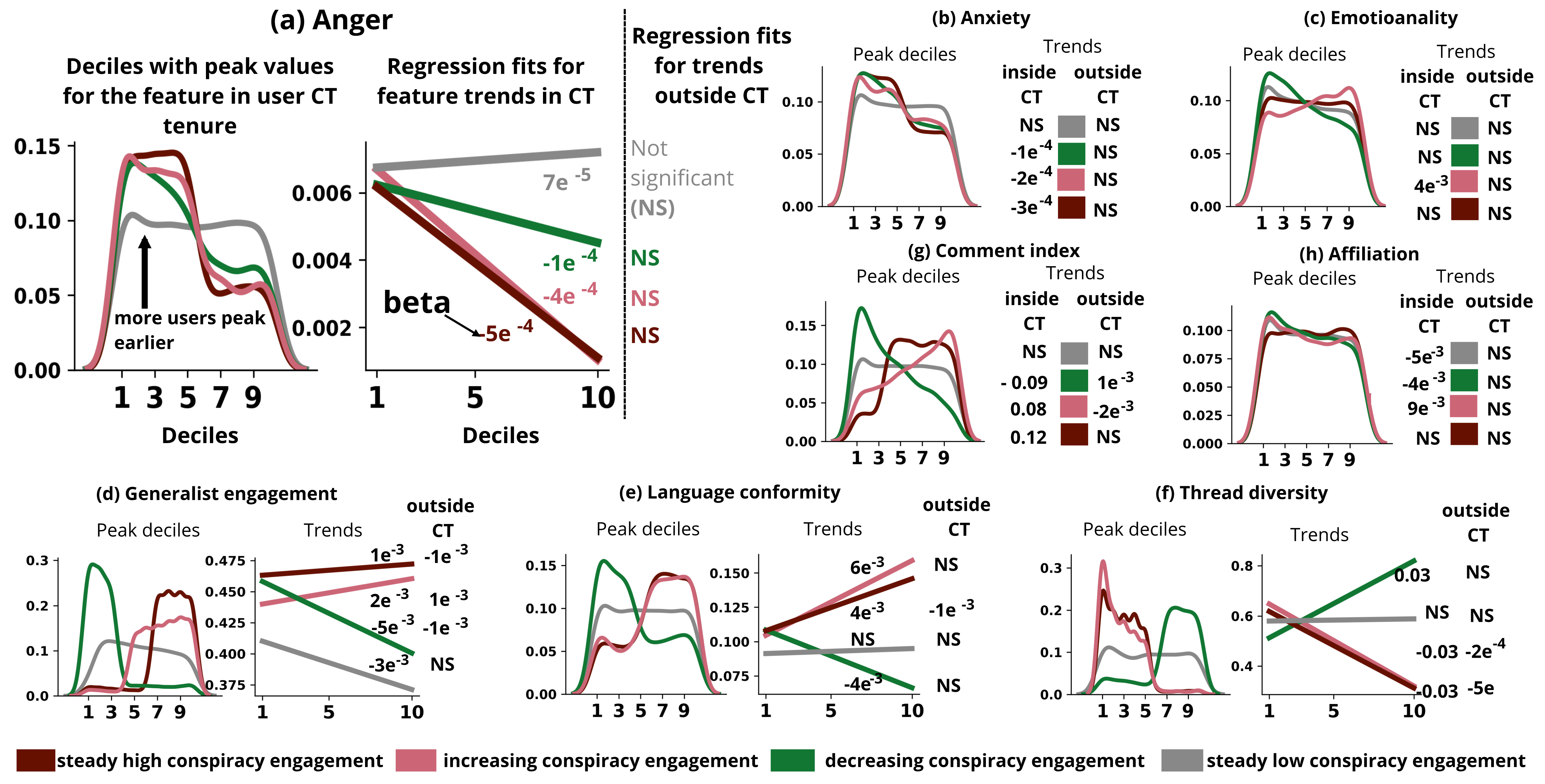}
    \caption{Figure presenting the peak deciles distributions and linear regression trends grouped by the conspiracy engagement pathways. In (a) we present an enlarged view of a typical result showing deciles with peak values and trends inside and outside of CT subreddits. 
    For all features, peak decile distributions represent the density plots for deciles at which the users attain the highest feature value. For example, users on all pathways show highest values for anger in earlier deciles (subfigure a). Trend in each subplot represents the linear regression fit for various pathways over deciles. For every line we denote the $\beta$ coefficient if the trend is significant. Non significant trends ($p >0.05$) are denoted with NS. We show trend coefficients for feature calculated both, inside and outside CT subreddits. Due to space limitations we show the actual trend lines only for 4 features.}
    \label{fig:featuremap}
\end{figure*}

\begin{figure*}[t]
    \centering
    \includegraphics[width=0.99\textwidth]{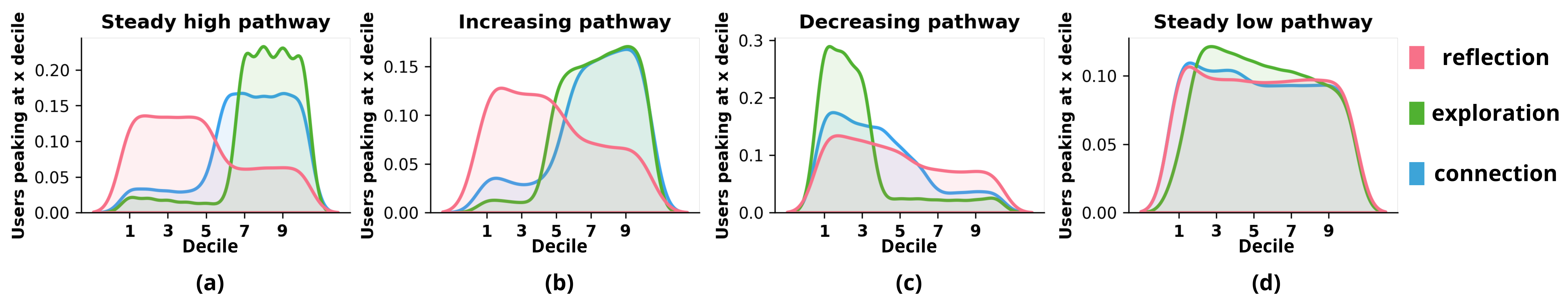}
    \caption{Figure outlining peak deciles for features in various radicalization phases for (a) \textit{steady high} (b) \textit{increasing} (c) \textit{decreasing} and (d) \textit{steady low} pathways. Peak in earlier deciles means that users attain highest values for that phase, early on. 
    }
    \label{fig:phase_progress}
\end{figure*}

\subsubsection{Language conformity:}
Cohesion or conformity with one's social group is a fundamental requirement in the process of radicalization \cite{crossett2010radicalization}. Especially in CT discussions, groups conform by establishing shared interpretations of reality around them \cite{butter2020routledge}. 
This process of  interpreting reality, or meaning-making, often manifests into the insider language used by conspiracy groups \cite{leone2017fundamentalism}. 
Hence, we measure language conformity by assessing how much of the subreddit's characteristic language does the user use.  

To measure a user's language conformity, 
we consider each subreddit the user contributes in as her ``social group'' and calculate the overlap between the language used by the user and that subreddit's characteristic language. To understand the characteristic language for each subreddit, we utilize Sparse Additive Generative models (SAGE) \cite{eisenstein2011sparse} that uses a regularized log-odds ratio to compare word distributions across various text corpora. 
We compare the word distributions in the text corpus of each subreddit with that of all other subreddits using SAGE. As a result, for each subreddit we obtain a lexicon of 1000 words that distinctively represent the language used in that subreddit. Table \ref{tab:lexconform} displays example lexicon words for various subreddits showing how SAGE is able to effectively capture the language specific to the discussions in each subreddit. 
Finally, we calculate a user's \textbf{language conformity} in a subreddit $s$ as the intersection of words used by the user in subreddit $s$ with the SAGE lexicon of the subreddit, normalized by total word count used by the user in $s$.


\subsubsection{Group connections:}
Interactions within small groups of like minded people can help in creating unambiguous shared narrative of events that is instrumental to the process of radicalization \cite{reedy2013terrorism}. Discussions within small groups can also limit the number of diverse opinions and information users might get exposed to, thus contributing to what researchers call as ``crippled epistemeology'' \cite{vermeule2009conspiracy}. 
Hence, we characterize users's small group interactions by analyzing the diversity of audience and repeated contributions in comment threads populated by users. 


First, we measure the diversity of audience in threads, or \textbf{thread diversity}, by comparing number of unique contributors to the total number of comments in a thread. 
While calculating the thread diversity ratio, we remove all contributions made by the subject user so as to not bias the ratio calculation by the user's contributions. Low \textit{thread diversity} would indicate that users engage in comment threads where limited number of other users contribute large number of comments. Second, to measure a user's involvement in grouped discussions, we calculate \textbf{comment rank}---the number of times a user repeatedly contributes in the same thread. 
Finally, we also measure group connections by analyzing how users express \textbf{affiliation} to their groups using LIWC's affiliation lexicon.

\subsection{Analyzing Evolution of Radicalization Phases}
\subsubsection{\textbf{1. Linear regression fits for trends: }} For each user on each conspiracy engagement pathway, we have ten values of all features corresponding to every decile. To understand how  features evolve over time, we fit a linear regression line with deciles as the independent variable and the feature value as the dependent variable. The magnitude of the fit coefficient ($\beta$) tells us the degree of increase and decrease over time and the p-value indicates whether the fit is significant. We display all trends lines and coefficients inside the 171 CT subreddits in Fig. \ref{fig:featuremap} along with the coefficients for trends \emph{outside} of conspiracy subreddits as well. Trend coefficients and significance outside CT show whether the trends we observe inside are specific to conspiracy discussion.

\subsubsection{\textbf{2. Users' conspiracy tenure with peak feature values: }}
We characterize phases of radicalization that, in theory, take effect one after another. Hence we next analyze how soon after CT joining, users attain peak values for features. For every user, we pick the decile with highest feature value and plot the density distribution of peak decile for all users grouped by conspiracy pathway. Density peak in early deciles would mean that users attain highest value for that feature immediately after initial participation in conspiracy discussions. 
All density plots are based on users' activity inside CT subreddits.  

\subsubsection{\textbf{3. Users' conspiracy tenure with peak phase features: }}
While the previous analysis displays how individual features peak across decile, here we group the feature belonging to same radicalization phases and plot similar peak density plots for each pathway (Fig. \ref{fig:phase_progress}). For example, Fig. \ref{fig:phase_progress} (a) indicates that for \textit{steady high} pathway, \textbf{R}eflection features peak immediately after initial conspiracy participation whereas \textbf{E}xploration and \textbf{C}onnection peak later in the CT journey. Visualizing this phase progression can inform whether users develop RECRO phases successively in time.

\section{Conspiracy Radicalization: Results}
\label{sec:rad_results}
\subsubsection{How do users on display markers of reflection phase?}
We characterized the \textbf{R}eflection phase using anger (Fig. \ref{fig:featuremap} (a)), anxiety ((Fig. \ref{fig:featuremap} (b)) and emotionality (Fig. \ref{fig:featuremap} (c)), expressed inside and outside of CT subreddits. Overall we find that use of language related to anger and anxiety decreases over time for users on all pathways. There are no significant trends for emotionality inside or outside of CT, except for users on the \textit{increasing} pathway who show increasing emotionality inside CT over time ($\beta$ = $4e^{-3}$). Earlier peaks in \textbf{R}eflection features may indicate what \citet{Neo2016AnRadicalisation} describes as ``cognitive opening'' where individuals turn to internet to express their grievances and vulnerabilities. Do all users advance to subsequent phases of radicalization after the cognitive opening? To find out, we next analyze the results of the \textbf{E}xploration phase. 

\subsubsection{How do users explore Reddit's conspiracy world?}
To characterize the \textbf{E}xploration phase, we investigate how users develop conspiracy worldview through generalist engagement (Fig. \ref{fig:featuremap} (d)). Higher generalist engagement would indicate that users engage in general CT discussions, thus developing monological worldview. We find that users on \textit{steady high} ($\beta$ = $1e^{-3}$), and \textit{increasing} pathway ($\beta$ = $2e^{-3}$) increasingly participate in the generalist CT subreddits. Interestingly, \textit{steady high} users show \emph{reduced} generalist engagement outside of CT subreddits ($\beta$ = $-1e^{-3}$). 
Conversely, users on \textit{decreasing} pathway increasingly contribute in specialist CT discussions ($\beta$ = $-5e^{-3}$) and have highest generalist engagement only in the earlier deciles. 
We ran an additional robustness check to ensure that this result is not an artifact of correlation between \textit{conspiracy similarity scale} and \textit{generality scale}.
Overall, we find that users with \textit{steady high} and \textit{increasing} CT engagement may also adhere to monological conspiracy worldview by increasingly participating in general conspiracy discussion subreddits.

\subsubsection{How do users make connections inside conspiracy communities?}
We measure group bonding through language conformity (Fig. \ref{fig:featuremap} (e)), audience diversity in threads (Fig. \ref{fig:featuremap} (f)), repeated comments (Fig. \ref{fig:featuremap} (g)) in threads, and affiliation related language (Fig. \ref{fig:featuremap} (h)). Overall, we find that users on \textit{steady high} ($\beta$ = $4e^{-3}$) and \textit{increasing} ($\beta$ = $6e^{-3}$) pathways develop high language conformity with CT subreddits. However, users on \textit{steady high} pathways show reduced language conformity outside of CT subreddits ($\beta$ = $-1e^{-3}$). Interestingly, users on \textit{decreasing} pathway ($\beta$ = $-4e^{-3}$), despite exhibiting early high engagement, never develop as high language conformity with CT subreddits in comparison to the other cohorts. Hence, early lexical conformity could be one of the important precursor of sustained CT engagement. Users with \textit{steady high} engagement also participate repeatedly ($\beta$=0.12) in smaller discussion groups with less audience diversity ($\beta$ = -0.03). Users on \textit{increasing} pathways also show similar trends. These results suggest that users on \textit{steady high} and \textit{increasing} pathways repeatedly show engagement in discussions with less diverse user base inside CT subreddits.   

\subsubsection{How does conspiracy radicalization evolve?}
Fig. \ref{fig:phase_progress} shows the deciles in which users attain peak feature values in different phases. We observe that in \textit{steady high} (Fig. \ref{fig:phase_progress} (a))  and \textit{increasing} (Fig. \ref{fig:phase_progress}(b)) pathways, users show higher feature values in the \textbf{R}eflection phase right after starting CT participation and develop high \textbf{E}xploration and \textbf{C}onnection feature values in later deciles. This may suggest that the internet-mediated conspiracy radicalization does evolve through different phases over time. Interestingly, users on \textit{decreasing} pathway show high feature values for all phases only early on, while for users on \textit{steady low} pathway, there is no discernible peak for these phases. We discuss the implications of these results in the discussion section.

\section{Discussion and Limitations}
In this paper, we present the first, large-scale modeling of long term engagement and radicalization in the online conspiracy communities using a longitudinal digital trace data of 36K Reddit users.
Below we discuss how our findings may impact the understanding of online CT participation and motivate further research.

\subsubsection{Monologicality as a varying process: }
While not all conspiracy believers adopt a general conspiratorial worldview as the primary sensemaking device \cite{franks2017beyond}, we find that two groups of users (those on \textit{steady high} and \textit{increasing} engagement trajectories) do contribute prominently in general conspiracy subreddits that host all types and topics of conspiracies. Are users on \textit{steady high} and \textit{increasing} pathways predisposed to monological thinking? The original theoretical proposition by \citet{goertzel1994belief} describes monologicality as a stable cognitive style, trait or disposition. However, looking specifically at users on \textit{increasing} pathway, the discussion spaces they engage in become more generalist over time. This suggests that monological conspiracy worldviews can develop over time. In fact, our quantitative results align with qualitative observation of \citet{franks2017beyond} depicting monologicality as a variable \emph{endpoint} of various social processes rather than a cognitive predisposition. 
In particular, counter to the popular rabbit-hole metaphor, individuals who show signs of radicalization do not seem to narrow their interests down to fringe theories. Instead, such individuals adopt venues of generalist CT discussions together with their idiosyncratic lingo. This observation, on the one hand, purports a parallelism between monological worldviews and radicalization. On the other, it begs the question of what types of online discussion environments harbor the potential for mobilization of radicalized individuals: topically and socially fringe spaces that may host extreme ideas, or comparatively mainstream spaces that afford perception biases of false consensus?





\subsubsection{Resolution and Operational Phases: }
We offer a systematic characterization for how radicalization in online CT discussions progresses, finding that users on \textit{steady high} and \textit{increasing} pathways do progress through these first three RECRO phases during their conspiracy tenure. The theoretical model suggests that users may then enter a resolution and operational phase \cite{Neo2016AnRadicalisation}. The resolution phase describes a period in which individuals gain momentum to convert their radical beliefs into action. While not all users that internalize radical beliefs actually act on them, those who do, act from a biased perspective solidified during the earlier stages \cite{Neo2016AnRadicalisation}. Simply put, users who go through all three, \textbf{R}eflection, \textbf{E}xploration and \textbf{C}onnection phases over time, change not just their beliefs but also their behaviors. Finally, the operational phase indicates a period in which individuals get mentally or physically prepared to commit acts that advance their radical objectives in the real world. They may influence others or actively look for openings to form plans of physical actions \cite{Neo2016AnRadicalisation}. 



\subsubsection{Recovery from online conspiracies: }
Perhaps one of the most unexplored, yet highly impactful research directions is to understand why and how users disengage from online conspiracy discussions. In this paper we provide quantitative evidence that a significant group of users do gradually decrease their participation in online conspiracy discussions while displaying measurably distinctive behaviors compared to \textit{steady high} and \textit{increasing} trajectories. For example, users on \textit{decreasing} pathway do not develop lexical conformity with conspiracy subreddits, engage in discussions containing diverse contributors and significantly reduce affiliation-related language in conspiracy subreddits. These observations provide valuable insights for understanding the process of recovery from conspiracy engagement. Our results provide ground to investigate conspiracy believers that do not adopt a monological conspiracy worldview separately from those who do. By understanding the difference between these two types of online conspiracy beliefs, our findings can help refocus research efforts on communities that have higher chances of advancing conspiracy radicalization. Conversely, the methods detailed in this paper can help in identifying users on \textit{decreasing} pathways who might be more receptive of cross-cutting narratives, and in transforming insights gained by studying them into design interventions, to counter the spread of disinformation and conspiracy theory radicalization. 


\section{Limitations and Future Work}

Our work has some limitations that should be acknowledged. \hl{First, we characterize the conspiracy engagement trajectories by using the contribution volumes of users, a measure commonly used as a strong latent proxy for user engagement \mbox{\cite{Hamilton2017LoyaltyCommunities}}. While this affords studying the complete evolution of contributions in the CT communities, analyzing contribution volume alone can limit the interpretability of quality of contributions. For example, it is possible that some users may be contributing troll posts while keeping the same contribution volume as others. A more nuanced measure of CT engagement could involve analyzing text and context of the user contributions.}
Second, this work offers empirical insights on how users escalate through the formative phases of radicalization. Yet, it would be crucial to unpack when and how this potential is turned into action in the \textbf{R}esolution and \textbf{O}perational phases. Our work provides a framework for experimental designs in this direction. 
\hl{Next, our characterization of CT engagement trajectories relies on subreddit contributions. We found that five subreddits higher up in the \textit{conspiracy similarity scale} were banned before 2020, which could have potentially affected the CT disengagement of some users. Currently, the literature examining the effect of subreddit bans on user engagement poses mixed results claiming that the bans are \mbox{\cite{Chandrasekharan2017,thomas2021behavior}} or aren't \mbox{\cite{habib2019act}} effective in specific cases they study. While the banned conspiratorial subreddits in our dataset made up for less than 0.2\% of the dataset volume, a more individualized investigation of the authors who prominently contributed in banned communities could reveal whether the subreddit bans actually affected the users' disengagement. }
Moreover, this paper observes the radicalization process from aggregated user activity. Qualitative analyses of CT narratives and of how those change across radicalization phases, should complement our work and provide a fuller understanding of the phenomenon. Also, this paper offers findings that are correlational in nature. Causal models could reveal more nuanced relationship between conspiracy engagement pathways and the radicalization process. 

\section{Additional Robustness Checks}
\label{sec:robustckech}

\noindent \textbf{1. User selection based on {\small \tt r/conspiracy}: } To characterize users' conspiracy radicalization process we select users that make at least 20 comments in {\small \tt r/conspiracy} over at least 1 year of time. How is our user selection affected by this {\small \tt r/conspiracy} constraint? To understand, we calculate the proportion of users' activity in other conspiracy related subreddits \emph{before} their first comment in {\small \tt r/conspiracy}. We find that 1,689 (4\%) of the users have more than 10\% of their total activity in other conspiracy subreddits. We remove the 1,689 users from entire analysis to ensure that {\small \tt r/conspiracy} is a common starting point into Reddit's conspiracy world for the users in our study. 

\noindent \textbf{2. Coverage for \textit{conspiracy similarity scale}: } Our \textit{conspiracy similarity scale} has conspiracy similarity value for only 4K subreddits. Does this mean we are missing out on modeling a large chunk of user activity while extracting conspiracy engagement pathways? We performed activity coverage analysis and found that just considering 4K subreddits in \textit{conspiracy similarity scale}, 90\% of authors have more than 80\% coverage of their total post-conspiracy joining Reddit activity. We removed the rest of the 10\% authors to ensure large coverage for all studied users.

\noindent \textbf{3. Correlation between conspiracy similarity and generality scale: } As we compare the generalist engagement for users on different conspiracy pathways, we wanted to ensure that two scales operating underneath are not correlated. Meaning, we wanted to check whether high conspiracy engagement inherently result in high generalist engagement due to our operationalizations. We performed Spearman rank correlation between subreddit rankings of the two scales and found only a weak correlation of 0.23. 

\noindent  \hl{\textbf{4. Effect of user removals: } To maintain the integrity of analysis across all users, we split the user activity in 10 deciles of equal contribution volumes. Meaning, all users have measurable activity in all of the deciles of their Reddit trajectory. This ensures that if a user gets banned, all activity before the ban will be studied across 10 deciles, indicating that user bans could not explain the observed disengagement in the user's trajectory prior to the ban.}  

\noindent  \hl{\textbf{5. Effect of subreddit bans: } To understand how subreddit bans could affect the disengagement, we compiled a list of over 3000 banned subreddits. Specifically, we sourced the list of banned subreddits from {\small \tt r/reclassified} subreddit which maintains an up-to-date list of banned subreddits. Since, our dataset contains user activities prior to 2020, we first identified subreddits that were banned prior to 2020. Out of the 4,756 subreddits in the dataset, 21 were banned prior to 2020. Out of the 21 banned subreddits, 5 subreddits lie in top 500 subreddits on the \textit{conspiracy similarity scale}, potentially affecting the engagement and disengagement trajectories of users. Before bans, the above mentioned five subreddits produced the following percent volume in the dataset---{\small \tt r/greatawakening} $(0.1\%)$, {\small \tt r/altright} $(0.01\%)$, {\small \tt r/uncensorednews} $(0.008\%)$,  {\small \tt r/911truth} $(0.003\%)$, {\small \tt r/sjwhate} $(0.001\%)$. Together, these subreddits make up for less than $0.2$\% data before getting banned.} 

\section{Conclusions}
In this paper we investigate the association between online CT discussion engagement and radicalization. Through an ensemble of computationally derived features backed by theoretical models, we observe three radicalization phases---\textbf{R}eflection, \textbf{E}xploration, \textbf{C}onnection---across four conspiracy engagement pathways. We find that high or increasing engagement in CT discussions online is also associated with successive phases of online radicalization symbolizing psychological predisposition, adoption of alternate worldviews and social bonds with others in CT communities. Conversely, users with decreasing engagement show qualitatively different CT interest compared to other users, limiting their CT discussion tenure to specialized conspiracy topics. Our results have implications in understanding the conspiracy recovery process.

\section{Ethics Statement}

\hl{We refer to the AAAI code of conduct and ethics guidelines that mention stakeholders, harm, privacy and confidentiality dimensions of ethical research and conduct. First, we acknowledge that all people, especially social media users and social computing researchers are stakeholders in this research. With this paper, we intend to contribute insights that can be considered while building safer online spaces for all. Furthermore, given that this study is retrospective and involves no interaction with the studied population, we do not anticipate any direct harm resulting from this research. We take proactive steps to preserve user privacy. Specifically, by presenting results aggregated over thousands of users and by intentionally not reporting any exact quotes made by Reddit users, we reduce the risk of re-identification. Finally, throughout this study, we analyze non-confidential Reddit data that is available in the public domain, collected through the publicly accessible Reddit Pushshift API \mbox{\cite{baumgartner2020pushshift}}. Yet, given the potential stigma associated with participating in CT discussions, we do not release any raw user data from this study.}

\section{Acknowledgments}
We thank the members of Social Computing and Algorithmic Experiences (SCALE) Lab at University of Washington for their valuable feedback on this work. This research was supported by Office of Naval Research (ONR-YIP \#N00014-21-1-2748), a US Navy/DOD Minerva (\#N00014-21-1-4001), and an NSF grant IIS (\#2041068).


\bibliography{mattiarefs,references}

\end{document}